\documentclass[a4paper,11pt]{article}  
  
\usepackage{amsfonts,amssymb}  
\usepackage{theorem}  

\def\G{\Gamma}    
\def\GG{{\rm I}\!\Gamma}

\begin{document}     

\rightline{MPI-PhT/2003-22}
\rightline{MPP-2003-71}
\vskip 1.6 truecm

\Large
\bf
\centerline{Higher-order non-symmetric counterterms}
\centerline{in pure Yang-Mills theory}

\normalsize
\rm

\vskip 0.5 truecm
\large
\centerline{Andrea Quadri\footnote{\tt quadri@mppmu.mpg.de}}

\vskip 0.5 truecm
\normalsize
\centerline{Max-Planck-Institut f\"ur Physik}
\centerline{(Werner-Heisenberg-Institut)}
\centerline{F\"ohringer Ring, 6 - D80805 M\"unchen, Germany}

\vskip 0.5  truecm
\normalsize
\bf
\centerline{Abstract}

\rm
{\small
\begin{quotation}
We analyze the restoration of the Slavnov-Taylor (ST) identities for pure
massless Yang-Mills theory in the Landau gauge within
the BPHZL renormalization scheme.
The Zimmermann-Lowenstein IR regulator $M(s-1)$ is introduced
via a suitable BRST doublet, thus preserving the nilpotency
of the BRST differential.
We explicitly obtain the most general
form of  the action-like part of the symmetric regularized action $\GG_s$,
$s <1$
obeying the ST identities and all other
relevant symmetries of the model, to all orders
in the loop expansion, and show that
non-symmetric counterterms arise in $\GG_s$ 
starting from the second order
in the loop expansion,
unless a special choice of normalization conditions is done.
We give
a cohomological characterization of the 
fulfillment of BPHZL IR power-counting criterion,
guaranteeing the existence of the physical limit $s \rightarrow 1$.

The technique analyzed in this paper is needed in the
study of the restoration of the ST identities for 
those models, like the MSSM, where massless particles 
are present and no invariant regularization scheme is known to preserve
all the relevant ST identities of the theory.

\end{quotation}
}

\vskip 0.7 cm

\begin{flushleft}
PACS Code: 11.10.Gh - 11.15.Bt.\\
Keywords: Renormalization, BRST Symmetry.
\end{flushleft}

\newpage

\section{Introduction}

In a preceding paper \cite{Quadri:2003ui}  
 the quantum restoration of the Slavnov-Taylor (ST) identities
for anomaly-free gauge theories
has been shown to be equivalent,
in the absence
of IR problems,
 to the recursive parameterization of the 
action-like part of the symmetric quantum effective action 
$$\GG=\sum_{j=0}^\infty \GG^{(j)} $$
in terms of suitable ST functionals, associated to the cohomology
classes of the classical linearized ST operator ${\cal S}_0$.
In the above equation  $\GG^{(j)}$ is the coefficient of order $j$ in the $\hbar$-expansion of $\GG$. It has also been shown \cite{Quadri:2003ui}
that even for models where a regularization-invariant
scheme\footnote{~ For instance dimensional regularization \cite{dim_reg}
for non-chiral gauge models or the modified subtraction prescription
given in \cite{tonin} for some chiral non-supersymmetric models.}
exists, at orders higher than one non-symmetric terms can enter
in $t^4 \GG^{(j)}$, $j \geq 2$, as a consequence of the
bilinear form of the ST identities. In the case discussed
in \cite{Quadri:2003ui} these non-invariant terms
can be put equal to zero only by a special choice of normalization conditions.

The method developed in \cite{Quadri:2003ui}, as well
as the techniques proposed in \cite{PLBorig}-\cite{Grassi:2001zz},
allows to recursively construct the symmetric 1-PI
Green functions, fulfilling the relevant
ST identities, without explicitly computing the
ST breaking terms which can appear at the regularized level.

In comparison with the methods based on the explicit recursive
evaluation of the ST breaking terms and of the corresponding
finite counterterms designed to recover 
them \cite{Martin:1999cc}-\cite{Hollik:1999xh},
the techniques aiming at the direct
restoration of the ST identities \cite{Quadri:2003ui}-\cite{Grassi:2001zz}
have the advantage to be regularization-scheme independent 
and to reduce the amount of
computations needed in order to obtain the correct symmetric
1-PI Green functions of the theory.

If the expansion of the action-like part of the symmetric vertex
functional is performed on a basis of Lorentz-scalar
monomials in the fields, the antifields and their derivatives,
a set of consistency conditions \cite{PLBorig,Ferrari:1998jy}
among the superficially convergent $n$-th order 1-PI Feynman amplitudes
and known lower orders 1-PI vertex functions
appears, reflecting
the nilpotency of the classical linearized ST operator
${\cal S}_0$ \cite{PLBorig}. These consistency conditions must be considered
while recursively solving the linear system 
whose solutions are the coefficients of the action-like monomials 
entering into the expansion 
of the correct symmetric 1-PI quantum effective action
\cite{PLBorig}. They become extremely involved for supersymmetric models.
On the contrary, when the symmetric 1-PI Green functions are
parameterized in terms of the ST functionals, according to the
prescription given in \cite{Quadri:2003ui}, these consistency
conditions are automatically taken into account.

\medskip
We remark that whenever no invariant regularization scheme is known,
the regularized Green functions do break the ST identities, and
therefore under these circumstances
the explicit computation of the finite restoring counterterms, required
to fulfill the relevant ST identities, cannot be avoided.

This is the case for instance 
of supersymmetric models like 
the Minimal Supersymmetric Standard Model (MSSM), for which 
no invariant regularization scheme is known to fulfill 
all the relevant symmetries of the theory,
due to the presence of the $\gamma^5$ matrix and of the 
completely antisymmetric tensor.

The complexity 
of the MSSM makes it very difficult to carry out 
first the recursive 
explicit computation of the ST breaking terms at 
the regularized level and then to obtain the
finite counterterms designed to recover them.
Therefore it would be very useful to be able to directly
reconstruct the symmetric 1-PI Green functions.
Unfortunately the procedure of \cite{Quadri:2003ui} 
can only be applied
in those cases where the relevant 1-PI Green functions
can be Taylor-expanded around zero momentum.
As a consequence this method, as well as the one analyzed in
\cite{PLBorig}, cannot be used to deal with
theories where massless particles are present. 

\medskip
The purpose of this paper is to extend the method
discussed in \cite{Quadri:2003ui} to those
theories where massless particles appear.
The technique analyzed in the present paper will allow
to deal with the 
Standard Model (SM) and the MSSM.
This in turn provides 
the last building block required 
in order to achieve the complete characterization,
 within the framework of the 
direct restoration of the ST identities for anomaly-free gauge
theories,
of the symmetric 1-PI Green functions for the SM and especially
of the MSSM, to all orders in the loop expansion.

\medskip
We will illustrate this extension in the case of pure 
massless Yang-Mills theory in the Landau gauge. 
This example is simple enough not to obscure the essential features
of the method, while retaining all the main properties
which make it useful in the study of the SM and of the MSSM.
We will obtain the explicit form, to all orders
in the loop expansion and for arbitray normalization conditions,
of the counterterms in the BPHZL scheme, required to fulfill
the ST identities of the model.

The  choice of the
Landau gauge simplifies the computations, but similar results 
can be derived for any Lorentz-covariant gauge.
We choose to work within the BPHZL regularization scheme
\cite{Bogolyubov:nc}-\cite{Zimmermann:1969jj},\cite{Lowenstein:ug}, 
following the Lowenstein-Zimmermann prescription 
\cite{Lowenstein:1975ps}-\cite{Lowenstein:1975rg}
 to handle massless propagators. Therefore all massless
fields are assigned a mass 
\begin{eqnarray}
m^2 = M^2 (s-1)^2 \, , 
\label{ir}
\end{eqnarray}
where $s$ is
an auxiliary parameter ranging between $0$ and $1$. 

The relevant zero-momentum subtractions on 1-PI Green functions,
required by the BPHZL scheme, 
take place both at $s=0$ and at $s=1$ according to the 
prescriptions given in \cite{Lowenstein:1975ps}-\cite{Lowenstein:1975rg}. 
Both subtractions are needed
in order to guarantee the existence of the massless limit 
$s\rightarrow 1$.

We will perform the renormalization of the relevant
ST identities for the intermediate regularized symmetric quantum
effective action $\GG_s$, $0\leq s<1$, by combining a variant of the approach
first pioneered in \cite{Lowenstein:pd} 
with the use of a BRST-invariant
IR regulator introduced via a BRST-doublet \cite{Blasi:1995vt}.
This allows to maintain nilpotency of the full BRST differential
and moreover provides the tool to discuss the interplay between
the cohomological properties of ${\cal S}_0$ and the subtraction
prescriptions required to guarantee the existence of the 
massless limit $s \rightarrow 1$.

The use of the IR regulator in eq.(\ref{ir}) allows to perform
a Taylor expansion of the 1-PI Green functions generated
by $\GG_s$, $s<1$, around zero momentum 
\cite{Lowenstein:1975ps,Lowenstein:1975rg}.
$\GG_s$, $s<1$ is required to fulfill a suitable extended ST identity
where the IR regulator $m$ in eq.(\ref{ir}) enters via a BRST doublet.
Then we can explicitly derive the most general solution of the action-like part
$t^4 \GG_s$ of $\GG_s$, $s<1$ by making use of the results given in 
\cite{Quadri:2003ui}. 
Since $s<1$, 
$t^4 \GG_s$ is well-defined and is obtained 
by expanding $\GG_s$
into a sum of linearly independent Lorentz-scalar
monomials in the fields, the antifields and their derivatives and
by keeping those terms of degree at most  $4$ in the mass dimension.

We find that non-symmetric counterterms enter into $\GG_s^{(j)}$, $j \geq 2$,
unless a special choice of normalization conditions has been done for 
$\GG^{(k)}_s$, $k<j$. The coefficients of the non-symmetric
counterterms in $\GG_s^{(j)}$ can be parameterized
in terms of the coefficients $\lambda_1^{(k)}, \rho_1^{(k)}$ of
suitable invariant ST functionals appearing in $t^4 \GG^{(k)}_s$,
$k < j$. 

It turns out that the limit of $\GG_s$ for $s \rightarrow 1$ is well-defined
and free of IR singularities,
as a consequence of the fulfillment of the IR and UV power-counting criteria
stated in \cite{Lowenstein:ug,Lowenstein:1975rg}.
The fulfillment of the IR power-counting criterion, guaranteeing
the absence of zero-mass singularities in the limit
$s \rightarrow 1$, can be understood in terms of purely
cohomological properties of the classical linearized
ST operator ${\cal S}_0$.

This provides a novel cohomological interpretation of the 
IR power-counting criterion first 
introduced in \cite{Lowenstein:ug,Lowenstein:1975rg}.

The quantum effective action $\GG$ for pure
massless Yang-Mills model is finally obtained by
\begin{eqnarray}
\GG = \lim_{s\rightarrow 1} \GG_s \, .
\label{intro1}
\end{eqnarray}
Physical unitarity stems from the ST identities obeyed by $\GG$.

We find that the non-symmetric counterterms entering into $\GG^{(j)}_s$
at order $j \geq 2$ do not vanish in the limit $s \rightarrow 1$.
Hence they also affect $\GG^{(j)}$,
unless a special choice of normalization
conditions has been done for $\GG^{(k)}_s$, $k<j$.

We point out that no explicit computation of the ST breaking
terms at the regularized level is needed in this construction.

Finally we comment on the dependence of physical observables
on the coefficients $\lambda_1^{(k)}, \rho_1^{(k)}$ 
parameterizing the non-invariant counterterms. 
This shows some of the advantages provided by the ST parameterization
introduced in \cite{Quadri:2003ui} in discussing the physical
consequences of the non-invariant higher order counterterms.

\medskip
The plan of the paper is the following. In Sect.~\ref{sec2}
we discuss the ST identities for the model at hand and provide
the most general solution to the symmetric regularized
quantum effective action $\GG_s$, $s<1$ to all orders
in the loop expansion. 
We show how the BPHZL IR regulator originally proposed in
\cite{Lowenstein:1975ps} can be introduced under the form of a suitable BRST
doublet.
This in turn allows to derive the proper ST identities
to be fulfilled by $\GG_s$, $s<1$.
We analyze the appearance of non-invariant counterterms
in $\GG_s^{(j)}$, $j\geq 2$. We show that they do not disappear
unless a special choice
of normalization conditions is done for $\GG_s^{(k)}$, $k<j$.
We give a cohomological interpretation of the IR power-counting criterion
stated in \cite{Lowenstein:ug,Lowenstein:1975rg} and show
that it is related to the structure of the extended BRST differential,
thus establishing a connection between the BPHZL treatment of IR divergences
and cohomology.
In view of the fulfillment of the IR power-counting criterion, which
we prove to hold true on the basis of purely cohomological arguments,
the limit of $\GG_s$ for $s \rightarrow 1$ exists. Therefore we can identify
the fully renormalized quantum effective action $\GG$ as the limit
$s \rightarrow 1$ of $\GG_s$. We discuss this identification
in Sect.~\ref{sec3}, where we also comment on
the dependence of physical observables on 
the parameters controlling the non-invariant counterterms.
Finally conclusions are presented in Sect.~\ref{sec4}.

\section{Higher-order non-symmetric counterterms}\label{sec2}

The ST identities for the classical action $\G^{(0)}$ 
(see Appendix~\ref{appA}) of 
Yang-Mills theory in the Landau gauge with
an IR regulator $m$ introduced via a BRST doublet $(\bar \rho,m)$ 
are
\begin{eqnarray}
{\cal S}(\G^{(0)}) = (\G^{(0)},\G^{(0)}) 
+ m \frac{\partial \G^{(0)}}{\partial \bar \rho}
= 0 \, ,
\label{st1}
\end{eqnarray}
where the bracket $(\G^{(0)},\G^{(0)})$ is defined according to
\begin{eqnarray}
(X,Y) = \int d^4x \, \left (
\frac{\delta X}{\delta A_\mu^{a*}} \frac{\delta Y}{\delta A^{\mu a}}
+
\frac{\delta X}{\delta \omega^{a*}} \frac{\delta Y}{\delta \omega^{a}} 
 + \frac{\delta X}{\delta \bar \omega^{a*}}
   \frac{\delta Y}{\delta \bar \omega^a} \right ) \, .
\label{st1_bis}
\end{eqnarray}
The linearized classical ST operator ${\cal S}_0$, given by
\begin{eqnarray}
{\cal S}_0 & = & \int d^4x \, \left ( 
\frac{\delta \G^{(0)}}{\delta A_\mu^{a*}} \frac{\delta}{\delta A^{\mu a}}
+
\frac{\delta \G^{(0)}}{\delta A^{\mu a}} \frac{\delta}{\delta A_\mu^{a*}}
+
\frac{\delta \G^{(0)}}{\delta \omega^{a*}} \frac{\delta}{\delta \omega^{a}} 
+
\frac{\delta \G^{(0)}}{\delta \omega^{a}} \frac{\delta}{\delta \omega^{a*}} 
\right . \nonumber \\ 
&& 
\left . ~~~~~~~~~~~ + \frac{\delta \G^{(0)}}{\delta \bar \omega^{a*}}
                    \frac{\delta}{\delta \bar \omega^a} +
		    \frac{\delta \G^{(0)}}{\delta \bar \omega^a}
		    \frac{\delta}{\delta \bar \omega^{a*}}
\right ) 
+ m \frac{\partial}{\partial \bar \rho}
\, , 
\label{st2}
\end{eqnarray}
is nilpotent: ${\cal S}_0^2=0$.

The BRST partner $\bar \rho$ of the mass regulator $m$ can be reabsorbed
by the following antifield redefinition:
\begin{eqnarray}
A^{a*'}_\mu = A^{a*}_\mu - \bar \rho m A^a_\mu \, , ~~~
\omega^{a*'} = \omega^{a*} + \bar \rho m \bar \omega^a \, , ~~~
\bar \omega^{a*'} = \bar \omega^{a*} - \bar \rho m \omega^a 
\, . 
\label{st3}
\end{eqnarray}
This is a consequence of the fact that $(\bar \rho,m)$ are cohomologically
trivial, pairing into a BRST doublet. In the new variables 
in eq.(\ref{st3}) $\G^{(0)}$ becomes
\begin{eqnarray}
&& \!\!\!\!\!\!\!\!\!\!\!\!\!\!\!\!\!\!\!\!\!\! \G^{(0)} 
 =  \int d^4x \, \left \{ -\frac{1}{4g^2} G_{\mu\nu}^a G^{\mu\nu \, a}
- \bar \omega^a \partial_\mu
(D^\mu \omega)^a + B^a \partial A^a \right . \nonumber \\
&& \!\!\!\!\!\!\!\!\!\!\!\!\!\! \left . + A_\mu^{a*'} (D^\mu \omega)^a - \omega^{* a'} \frac{1}{2}f^{abc}
     \omega^b \omega^c + 
     \bar \omega^{a*'} B^a 
   + \frac{1}{2}m^2 (A_\mu^a)^2 + m^2 \bar \omega^a \omega^a 
   \right \} .
\label{st4}
\end{eqnarray}
The classical action $\G^{(0)}$ in eq.(\ref{st4}) 
obeys a set of additional symmetries:
\begin{eqnarray}
\frac{\partial \G^{(0)}}{\partial \bar \rho} = 0 \, ,
\label{aux1}
\end{eqnarray}
the $B$-equation
\begin{eqnarray}
\frac{\delta \G^{(0)}}{\delta B^a} = \partial A^a 
+  \bar \omega^{a*'}
\, , 
\label{aux2}
\end{eqnarray}
the ghost equation
\begin{eqnarray}
\frac{\delta \G^{(0)}}{\delta \bar \omega^a} + 
\partial^\mu \frac{\delta \G^{(0)}}{\delta A^{\mu a*'}} = m^2 \omega^a \, , 
\label{aux3}
\end{eqnarray}
and the anti-ghost equation
\begin{eqnarray}
&& \int d^4x \, \left ( \frac{\delta \G^{(0)}}{\delta \omega^a}
-f^{abc} \bar \omega^b \frac{\delta \G^{(0)}}{\delta B^c} \right )
\nonumber \\
&& ~~~~~ = \int d^4x \, \left ( m^2 \bar \omega^a 
- f^{abc} A_\mu^{b*'} A^{\mu c}
+ f^{abc} \omega^{*b'} \omega^c 
\right ) \, .
\label{aux4}
\end{eqnarray}
Notice that the R.H.S. of eqs.(\ref{aux2})-(\ref{aux4}) are linear
in the quantum fields.

\medskip

The ST identities in eq.(\ref{st1}) can be regarded 
from the point of view of the Batalin-Vilkovisky formalism
\cite{batalin} as the master equation \cite{batalin,Gomis:1994he}
for Yang-Mills theory in the presence of an IR regulator giving mass
to the gauge and ghost fields.
In eq.(\ref{st1}) $A^{a*}_\mu$, $\omega^{a*}$ and $\bar \omega^{a*}$
denote the antifields \cite{zj} coupled in $\G^{(0)}$ in eq.(\ref{mod1})
with the BRST variation respectively of the quantum fields
$A^a_\mu, \omega^a$ and $\bar \omega^a$.
The constant anticommuting parameter $\bar \rho$ pairs with the IR
regulator $m$ into a ${\cal S}_0$-doublet.

Since the relevant classical linearized ST operator ${\cal S}_0$ in 
eq.(\ref{st2}) is nilpotent, the cohomological analysis of the IR-regularized
Yang-Mills theory, whose classical action is given by 
$\G^{(0)}$ in eq.(\ref{mod1}), can be performed by making use of the methods
of Algebraic Renormalization  \cite{ps,Gomis:1994he,Barnich:2000zw}.

In particular, since the IR regulator $m$ forms a ${\cal S}_0$-doublet
together with $\bar \rho$, the cohomology of ${\cal S}_0$ is isomorphic
to that of the classical linearized ST operator of pure Yang-Mills theory
\cite{ps,Barnich:2000zw,Barnich:db,Barnich:mt}.

By exploiting this result it can be shown on purely algebraic grounds
\cite{ps,Barnich:2000zw} that the extended ST identities for the IR-regularized
Yang-Mills theory can be restored at the quantum level.
Therefore it is possible to construct a symmetric quantum effective
action $\G_m$ fulfilling the extended ST identities
\begin{eqnarray}
{\cal S}(\G_m) \equiv (\G_m,\G_m) + m \frac{\partial \G_m}{\partial \bar \rho} = 0
\label{regST}
\end{eqnarray}
to all orders in the loop expansion. The bracket in eq.(\ref{regST})
is defined by eq.(\ref{st1_bis}).
Nevertheless, this is not enough to guarantee
the existence of the massless limit $m \rightarrow 0$.
Indeed it may very well happen that the limit
\begin{eqnarray}
\G = \lim_{m \rightarrow 0} \G_m
\label{mass_limit}
\end{eqnarray}
is ill-defined, although $\G_m$ exists and fulfills eq.(\ref{regST}).

In order to discuss this point and to carry out properly the renormalization 
of the model
we choose to work within the BPHZL regularization scheme 
\cite{Bogolyubov:nc}-\cite{Zimmermann:1969jj},\cite{Lowenstein:ug}, 
by following the Lowenstein-Zimmermann prescription 
\cite{Lowenstein:1975ps}-\cite{Lowenstein:1975rg}
 to handle massless propagators.
 
For that purpose we identify
\begin{eqnarray}
m=M(s-1) \, ,
\label{ident}
\end{eqnarray}
where $0 \leq s \leq 1$ and $M$ is 
a constant with the dimension of a mass. The subtraction operator
$t_\gamma$ for a given divergent 1-PI graph or subgraph $\gamma$ involves
both a subtraction around $p=0,s=0$ and around $p=0,s=1$
\cite{Lowenstein:pd,Lowenstein:1975ps,Lowenstein:ug,Lowenstein:1975rg}:
\begin{eqnarray}
(1 - t_\gamma) = (1 -t^{\rho(\gamma)-1}_{p,s-1})(1-t^{\delta(\gamma)}_{p,s}) 
\, ,
\label{st5}
\end{eqnarray}
where $\rho(\gamma)$ is the IR subtraction degree  and
$\delta(\gamma)$ the UV subtraction degree for $\gamma$
\cite{Lowenstein:pd,Lowenstein:1975ps,Lowenstein:ug,Lowenstein:1975rg}.
We point out that both subtractions around $s=0$ 
and $s=1$ are needed in order to
guarantee the absence of IR singularities of the 1-PI Green functions
in the physical limit $s \rightarrow 1$ ($m \rightarrow 0$).

The assignments of UV dimension for the fields and external
sources, required to compute $\rho(\gamma)$ and $\delta(\gamma)$
for a given graph $\gamma$ involving the fields and the antifields
of the model, are as follows: 
$A_\mu^a, \omega^a, \bar \omega^a$ have UV dimension $1$, 
$B_a, A_\mu^{'a*}, \omega^{a*'}$ and $\bar \omega^{a*'}$
have UV dimension $2$.
The IR dimension coincides with the UV dimension.

Let us denote by $\GG_s$, $s <1$ the symmetric quantum effective action,
constructed according to the BPHZL subtraction prescription,
fulfilling the ST identities
\begin{eqnarray}
{\cal S}(\GG_s) =  \int d^4x \, \left (
\frac{\delta \GG_s}{\delta A_\mu^{a*}} \frac{\delta \GG_s}{\delta A^{\mu a}}
+
\frac{\delta \GG_s}{\delta \omega^{a*}} \frac{\delta \GG_s}{\delta \omega^{a}} 
 + \frac{\delta \GG_s}{\delta \bar \omega^{a*}}
   \frac{\delta \GG_s}{\delta \bar \omega^a} \right ) 
+ m \frac{\partial \GG_s}{\partial \bar \rho}
= 0 \, . 
\label{reg1}
\end{eqnarray}

During the renormalization procedure we will always keep $s<1$.
Only in the very end we will take the physical limit $s\rightarrow 1$.
We remark that for $s\neq 1$ the ST identities in eq.(\ref{reg1})
give rise to a violation of physical unitarity, due to the soft
breaking term
\begin{eqnarray}
M (s-1) \frac{\partial \GG_s}{\partial \bar \rho} \, .
\label{brk1}
\end{eqnarray}
This can be explicitly verified by using methods close
to the one discussed in \cite{Picariello:2001ri}. We recover
physical unitarity in the limit $s \rightarrow 1$.

It can be proven by using the standard methods discussed
e.g. in \cite{ps} that the functional identities
in eqs.(\ref{aux1})-(\ref{aux4}) 
can be restored at the quantum level. So we assume that they are
also fulfilled by the symmetric quantum effective action $\GG_s$:
\begin{eqnarray}
&& \frac{\partial \GG_s^{(j)}}{\partial \bar \rho} = 0 \, , ~~~~~
   \frac{\delta \GG_s^{(j)}}{\delta B^a} = 0 \, ,
~~~~~
  \frac{\delta \GG_s^{(j)}}{\delta \bar \omega^a} + 
  \partial^\mu \frac{\delta \GG_s^{(j)}}{\delta A^{a*'\mu}} = 
0 \, , 
\nonumber \\
&&
  \int d^4x \, \left ( \frac{\delta \GG_s^{(j)}}{\delta \omega^a}
-f^{abc} \bar \omega^b \frac{\delta \GG_s^{(j)}}{\delta B^c} \right )
= 0 \, , ~~~~~~~~~~~~~~~~~ j \geq 1 \, .
\label{aux5}
\end{eqnarray}
From the third equation in the first line of eq.(\ref{aux5})
we conclude that $\GG^{(j)}_s$, $j\geq 1$ depends
on $\bar \omega^a$ only through the combination
\begin{eqnarray}
\hat A^{*a'}_\mu = A^{a*'}_\mu + \partial_\mu \bar \omega^a \, .
\label{aux6}
\end{eqnarray}

\subsection{One-loop order}

At one-loop order the ST identities read
\begin{eqnarray}
{\cal S}_0 (\GG^{(1)}_s)=0 \, .
\label{l1}
\end{eqnarray}
The action-like part of the 
most general solution $\GG^{(1)}_s$ to eq.~(\ref{l1}), compatible with
the additional symmetries in eq.~(\ref{aux5}), 
is constrained to have the form 
\cite{ps,Barnich:mt,Barnich:2000zw,Quadri:2003ui}
\begin{eqnarray}
t^4 \GG^{(1)}_s & = & \lambda_1^{(1)} \int d^4x \, G_{\mu\nu}^a G^{\mu\nu}_a 
+ \rho_1^{(1)} {\cal S}_0 (\int d^4x \, \hat A^{a*'}_\mu A^a_\mu ) 
\, ,
 \label{ym23}
\end{eqnarray}
where $t^4$ is the projection operator on the sector
of dimension $\leq 4$ in the fields, the antifields and their derivatives.
$t^4 \GG^{(1)}_s$ exists since $s<1$. 

Let us comment on the R.H.S. of eq.(\ref{ym23}). At one-loop level
only ${\cal S}_0$-invariant terms appear in $t^4 \GG^{(1)}_s$.
Moreover, we notice that they are all IR-safe (all monomials
entering into the R.H.S. of eq.(\ref{ym23}) have IR degree
equal to $4$). This follows since
\begin{eqnarray}
{\cal S}_0 (\Phi^{*'}) = \frac{\delta \G^{(0)}_{m=0}}{\delta \Phi} 
\label{rel_not}
\end{eqnarray}
for $\Phi^{*'} = \hat A^{a*'}_\mu, \omega^{a*'}, \bar \omega^{a*'}$.
We will discuss this point further in the next subsection.

$\lambda_1^{(1)}, \rho_1^{(1)}$ are free parameters entering into
the solution, unconstrained by the ST identities and the additional
symmetries in eq.(\ref{aux5}). They can be fixed by providing
a choice of normalization conditions. As an example, one might
choose\footnote{The notation is as follows. We expand $t^4 \GG^{(j)}_s$
into a sum of linearly independent, Lorentz-scalar action-like functionals
${\cal M}_j(x)$ in the fields, the antifields and their
derivatives, providing a basis for the space to which
$t^4 \GG_s^{(j)}$ belongs, and write accordingly 
$t^4 \GG_s^{(j)} = \sum_l \int d^4x \, \xi^{(j)}_l {\cal M}_l(x)$. $\xi^{(j)}_l$ is the coefficient of ${\cal M}_l(x)$
in this expansion.
As an example, 
$\xi^{(1)}_{G_{\mu\nu}^a G^{\mu\nu}_a}$ is  the coefficient
of $G_{\mu\nu}^a G^{\mu\nu}_a$ , 
$\xi^{(1)}_{\hat A^{a*'}_\mu \partial^\mu \omega^a}$
the coefficient of $\hat A^{a*'}_\mu \partial^\mu \omega^a$ in the
expansion of $t^4 \GG_s^{(1)}$.}
\begin{eqnarray}
\xi^{(1)}_{G_{\mu\nu}^a G^{\mu\nu}_a} = 0 \, , ~~~~
\xi^{(1)}_{\hat A^{a*'}_\mu \partial^\mu \omega^a} = 0 \, ,
\label{cn1}
\end{eqnarray}
yielding
\begin{eqnarray}
\lambda_1^{(1)} = 0 \, ,  ~~~~~ \rho_1^{(1)} = 0 \, .
\label{cn2}
\end{eqnarray}
In the following we will not restrict ourselves 
to a special choice of normalization
conditions, so we will keep $\lambda_1^{(1)}, \rho_1^{(1)}$ free.

\subsection{Higher orders}

At orders higher than one the ST identities read
\begin{eqnarray}
{\cal S}_0(\GG_s^{(n)}) = -\sum_{j=1}^{(n-1)} (\GG_s^{(n-j)},\GG_s^{(j)}) \, .
\label{cn3}
\end{eqnarray}
The brackets are given in eq.(\ref{st1_bis}). Eq.(\ref{cn3}) is an inhomogeneous
linear equation whose unknown is the action-like part of $\GG^{(n)}_s$.
The fact that the non-local terms from the R.H.S. of the above equation
cancel against the non-local contributions from the L.H.S. is a consequence
of the Quantum Action Principle \cite{QAP} and of the assumption
that the ST identities have been restored up to order $n-1$.
Therefore we can restrict ourselves to the local approximation
of $\GG^{(n)}_s$, the formal power series in the fields, the antifields
and their derivatives corresponding to the Taylor expansion of all
relevant 1-PI Green functions around zero momentum.
This is possible since $s<1$.

The solution $t^4 \GG^{(n)}_s$ to eq.(\ref{cn3}) is \cite{Quadri:2003ui}
\begin{eqnarray}
t^4 \GG_s^{(n)} & = & 
\lambda_1^{(n)} \int d^4x \, G_{\mu\nu}^a G^{\mu\nu}_a 
+ \rho_1^{(n)} {\cal S}_0 (\int d^4x \, \hat A^{a*'}_\mu A^a_\mu ) 
\nonumber \\
& & 
+ \int d^4x A_\mu^a \frac{\delta}{\delta A_\mu^a} 
\, \left [ 
\sum_{j=1}^{n-1} \rho_1^{(n-j)} \lambda_1^{(j)}  
\int d^4y \, G^{\rho\sigma b} G_{\rho\sigma}^b 
\right ]
\nonumber \\
& & 
+ \sum_{j=1}^{n-1} \rho_1^{(n-j)}\rho_1^{(j)} 
\int d^4x \, A_\mu^a \frac{\delta}{\delta A_\mu^a} 
{\cal S}_0 ( \int d^4y \, \hat A^{b*'}_\nu A^{\nu b} )
\nonumber \\
& & 
- \sum_{j=1}^{n-1} 
\rho_1^{(n-j)}\rho_1^{(j)} \Big (
 \int d^4x \, 
\frac{1}{g^2} ( \square A^d_\rho - \partial_\rho (\partial A)^d ) A^{\rho d}
\nonumber \\
& &
- \int d^4x \, 
\frac{2}{g^2} f^{dlm} A^l_\sigma (\partial_\rho A^m_\sigma -
					 \partial_\sigma A^m_\rho) A^d_\rho 
\nonumber \\
& & 
 + \int d^4x \, 
\frac{1}{4g^2} f^{vqk}f^{krd} A_\sigma^q A_\sigma^r 
A_\rho^v A_\rho^d 
\Big ) \, .
\nonumber \\
\label{ym37}
\end{eqnarray}
In contrast with one-loop level, non-symmetric counterterms enter 
in $t^4 \GG_s^{(n)}$. They appear due to the inhomogeneous term 
in the R.H.S. of eq.(\ref{cn3}), which depends on $\GG_s^{(k)}$, $k<j$.

In an arbitrary Lorentz-covariant gauge an additional invariant
appears in the R.H.S. of eq.(\ref{ym23}), given by
\begin{eqnarray}
\rho_2^{(1)} {\cal S}_0 \Big ( \int d^4x \, \omega^{a* '} \omega^a \Big ) \, .
\label{new_inv}
\end{eqnarray}
This in turn implies that the R.H.S. of eq.(\ref{cn3}) gets more involved,
forbidding the application of the elegant homotopy techniques 
\cite{Quadri:2003ui}
which in the case of the Landau gauge allow to easily solve
the inhomogeneous problem in eq.(\ref{cn3}).
As we will show in a moment, 
the IR properties of the theory and their relationship 
with the 
cohomology of the operator ${\cal S}_0$ do not depend
on the choice of the gauge.
Therefore we choose to restrict ourselves to the Landau gauge, for which
there exists the simple and compact form for the general
solution $t^4 \GG^{(n)}_s$ 
given by eq.(\ref{ym37}).

\medskip
The non-symmetric counter-terms, 
depending on the lower order contributions, 
disappear if one chooses to impose the following normalization
condition for $t^4\GG_s^{(j)}$:
\begin{eqnarray}
\rho_1^{(j)}=0 \, , ~~~~~ 
j=1,2,\dots,n-1
\label{ym38}
\end{eqnarray}
equivalent to
\begin{eqnarray}
\xi^{(j)}_{\hat A_\mu^{a*'}\partial^\mu \omega_a} = 0 \, , ~~~~
j=1,2,\dots,n-1 \, .
\label{ym39}
\end{eqnarray}
We might supplement it by the choice
\begin{eqnarray}
\lambda^{(j)}_1=0 \, , ~~~~~
j=1,2,\dots,n-1 \, ,
\label{bis}
\end{eqnarray}
equivalent to
\begin{eqnarray}
\xi^{(j)}_{G^a_{\mu\nu}G^{\mu\nu a}}=0 \, , ~~~~ 
j=1,2,\dots,n-1 \, .
\label{ter}
\end{eqnarray}
Eqs.~(\ref{ym39}) and (\ref{ter}) extend the one-loop normalization conditions in eq.(\ref{cn1}).

\medskip
One can verify that all monomials in the R.H.S. of
eq.(\ref{ym37}) are IR safe by eq.(\ref{rel_not}).
We wish to comment on the IR-safeness of the monomials
entering into $t^4 \GG^{(n)}_s$. 
In the present approach
this property stems from the fact that the classical linearized ST operator
${\cal S}_0$ in eq.(\ref{st2}) 
has in the primed variables in eq.(\ref{st3})
definite degree $+1$ with respect to the counting
operator of the fields, the antifields and their derivatives.
The existence of the antifield redefinition in eq.(\ref{st3}) is in
turn a consequence of the fact that $(\bar \rho,m = M(s-1))$ form a 
BRST doublet. 
Since this is true also in an arbitrary Lorentz-covariant gauge,
the above result extends to that case too.

In \cite{Lowenstein:ug,Lowenstein:1975rg} the IR
power-counting criteria were obtained by means of
convergence arguments only and they did not display
any relationship with the cohomological properties of
${\cal S}_0$. Their cohomological interpretation in terms
of the degree of ${\cal S}_0$, made possible by the identification
in eq.(\ref{ident}), now shows that they can be actually understood
on purely cohomological grounds.

\section{The limit $s \rightarrow 1$}\label{sec3}

In the previous section we have derived the most general
form of the action-like part of $\GG_s$, $s<1$ to all order in the loop
expansion. We have checked that all action-like terms in $\GG_s$, $s<1$
are IR-safe. UV convergence criteria are also satisfied.
This is a sufficient condition 
\cite{Lowenstein:pd,Lowenstein:1975ps,Lowenstein:ug,Lowenstein:1975rg}
to guarantee the existence of the limit $s \rightarrow 1$:
\begin{eqnarray}
\GG=\lim_{s\rightarrow 1} \GG_s \, .
\label{ym40}
\end{eqnarray}
In the limit $s \rightarrow 1$ ($m \rightarrow 0$)
the primed antifields in eq.(\ref{st3})
reduce to their unprimed counterparts. 
The ST identities obeyed by $\GG$ read
\begin{eqnarray}
{\cal S}(\GG) =  \int d^4x \, \left (
\frac{\delta \GG}{\delta A_\mu^{a*}} \frac{\delta \GG}{\delta A_\mu^{a}}
+
\frac{\delta \GG}{\delta \omega^{a*}} \frac{\delta \GG}{\delta \omega^{a}} 
 + \frac{\delta \GG}{\delta \bar \omega^{a*}}
   \frac{\delta \GG}{\delta \bar \omega^a} \right ) 
= 0 \, . 
\label{final_st}
\end{eqnarray}
The soft-breaking term in eq.(\ref{brk1}) has disappeared. This
ensures the physical unitarity of the model \cite{cf}-\cite{becchi1983}. 
The non-symmetric counterterms entering into $\GG^{(j)}_s$
at order $j \geq 2$ do not vanish in the limit $s \rightarrow 1$.
Hence they also appear in $\GG^{(j)}$,
unless the special choice of normalization
conditions in eq.(\ref{ym39}) has been done for $\GG^{(k)}_s$,
$k < j$.

\medskip
We wish to comment on the dependence of physical observables on 
the parameters $\rho_1^{(j)}$. 
$\rho_1^{(j)}$ enters in $\GG^{(j)}_s$ as the coefficient
of the ${\cal S}_0$-exact functional 
${\cal S}_0 (\int d^4x \, \hat A^{a*'}_\mu A^a_\mu )$. 
Hence physical observables should not depend on $\rho_1^{(j)}$.
The study of the dependence of the Green functions of local
BRST invariant operators %${\cal O}(x)$ 
on these parameters
can be carried out according to the standard procedure
\cite{ps,sibold}, relying on the extension of the BRST differential $s$
in such a way to incorporate $\rho_1^{(j)}$ into a BRST doublet
together with its partner $\theta_1^{(j)}$:
\begin{eqnarray}
s \rho_1^{(j)} = \theta_1^{(j)} \, , ~~~~ s \theta_1^{(j)} = 0 \, .
\label{ext1}
\end{eqnarray}
The corresponding ST identities yield for $\GG$
\begin{eqnarray}
{\cal S}'(\GG) = {\cal S}(\GG) + 
\sum_j \theta_1^{(j)} \frac{\partial \GG}{\partial \rho_1^{(j)}} = 0 \, ,
\label{ext2}
\end{eqnarray}
where ${\cal S}(\GG)$ is given in eq.(\ref{final_st}).
Upon taking the Legendre transform $W$ of $\GG$ with respect to the quantized
fields of the model the ST identities read for the connected generating
functional $W$:
\begin{eqnarray}
{\cal S}'(W)= -\int d^4x \, \left ( 
J^a_{\mu} \frac{\delta W}{\delta A_\mu^{a*}}
+ J^a_\omega \frac{\delta W}{\delta \omega^{a*}}
+ J^a_{\bar \omega} \frac{\delta W}{\delta \bar \omega^{a*}}
\right ) + 
\sum_j \theta_1^{(j)} \frac{\partial W}{\partial \rho_1^{(j)}} =0 \, .
\label{ext3}
\end{eqnarray}
In the above equation $J^a_\mu$ is the external source coupled to
$A_{a\mu}$, $J^a_\omega$ the source coupled to $\omega^a$ and
$J^a_{\bar \omega}$ the source coupled to $\bar \omega^a$.
Now we differentiate eq.(\ref{ext3}) with respect to $\theta_1^{(j)}$
and with respect to the sources $\beta_1(x_1), \dots ,\beta_n(x_n)$,
coupled to local BRST-invariant operators ${\cal O}_1(x_1), \dots
{\cal O}_n(x_n)$ and go on-shell ($J=\beta=\theta=0$). We obtain
\begin{eqnarray}
\left . \frac{\delta^{(n+1)} W}{\delta \rho_1^{(j)} \delta \beta_n(x_n)
\dots \delta \beta_1(x_1)} \right |_{o.s.} = 0 \, .
\label{ext4}
\end{eqnarray}
Therefore the Green functions of local BRST-invariant operators
are $\rho_1^{(j)}$-independent, as a consequence of the ST identities
in eq.(\ref{ext2}).
At the level of the 1-PI generating functional
this property is encoded into the Nielsen-like identity
\cite{Nielsen:fs,sibold,Kluberg-Stern:rs,Kluberg-Stern:1974xv}
obtained by differentiating eq.(\ref{ext2}) w.r.t. $\theta_1^{(j)}$:
\begin{eqnarray}
\frac{\partial \GG}{\partial \rho_1^{(j)}} = 
\hat {\cal S}_{\GG} \left ( \frac{\partial \GG}{\partial \theta_1^{(j)}}
\right ) \, ,
\label{niels1}
\end{eqnarray}
where
\begin{eqnarray}
&& \!\!\!\!\!\!\!\!\!\!\!\!\!\!\!\!\!\!\!\!\!\!\!\!\!\!\!\!\!\!\!\!\!\!\!
\hat {\cal S}_{\GG}  =  
\int d^4x \, \Big ( \frac{\delta \GG}{\delta A^{a*}_\mu}
                    \frac{\delta}{\delta A^a_\mu}
                  + \frac{\delta \GG}{\delta A^{a}_\mu}
                    \frac{\delta}{\delta A_\mu^{a*}} 
%                  \nonumber \\
%              & &    
                  + \frac{\delta \GG}{\delta \omega^{a*}}
                    \frac{\delta}{\delta \omega^a}
                  + \frac{\delta \GG}{\delta \omega^{a}}
                    \frac{\delta}{\delta \omega^{a*}} 
                  \nonumber \\
              & &    + \frac{\delta \GG}{\delta \bar \omega^{a*}}
                    \frac{\delta}{\delta \bar \omega^a}
                  + \frac{\delta \GG}{\delta \bar \omega^{a}}
                    \frac{\delta}{\delta \bar \omega^{a*}} \Big )
%                  \nonumber \\
%              & &    
             + \sum_k \theta_1^{(k)} \frac{\partial}{\partial \rho_1^{(k)}} \, .
\label{niels2}
\end{eqnarray}
The possible deformations of eq.(\ref{niels1}), compatible
with the Quantum Action Principle and nilpotency of 
$\hat{\cal S}_{\GG}$, can be studied 
by using the methods developed in 
\cite{sibold,Gambino:1999ai}. 
Under the assumption that the
ST identities in eq.(\ref{final_st}) hold true
(so that $\hat {\cal S}_{\GG}^2=0$)
the most general structure
of the renormalized Nielsen-like identity turns out to be
\begin{eqnarray}
\frac{\partial \GG}{\partial \rho_1^{(j)}} = (1 + \sigma^\rho)
\hat {\cal S}_{\GG} \left ( \frac{\partial \GG}{\partial \theta_1^{(j)}}
\right ) + \beta^\rho \frac{\partial}{\partial g} \GG
+ \sum_{\varphi} \gamma_\varphi^\rho {\cal N}_\varphi \GG \, .
\label{niels3}
\end{eqnarray}
In the above equation $\beta^\rho$ parameterizes the explicit
dependence of the coupling constant $g$ on $\rho_1^{(j)}$.
Such a dependence
might be induced by the choice of normalization conditions
for the physical parameter $g$ that explicitly depend on $\rho_1^{(j)}$.
$\varphi$ stands for any of the fields of the model and
${\cal N}_\varphi$ denotes the corresponding counting operator.
$\sigma^\rho$, $\beta^\rho$ and $\gamma_\varphi^\rho$ control the 
deformations of the Nielsen identity in eq.(\ref{niels1}).

It is possible to recursively choose the counterterms of the model,
order by order in the loop expansion, in such a way that
$$\sigma^\rho=0, ~~~~ \beta^\rho=0, ~~~~ \gamma_\varphi^\rho =0 \, .$$
In this case we recover eq.(\ref{niels1}).

However, one can relax the conditions on $\sigma^\rho$ and
$\gamma_\varphi^\rho$ while preserving %all the same 
the property
of the independence of physical observables of $\rho_1^{(j)}$.
Indeed, since the terms ${\cal N}_\varphi \GG$ are equivalent to the
insertion of ${\cal S}_{\GG}$-exact local operators \cite{ps,Gambino:1999ai},
they do not affect the dependence of physical Green functions on
$\rho_1^{(j)}$ \cite{Gambino:1999ai}. 
The same is true for the first term in the R.H.S. of eq.(\ref{niels3}).

Therefore the only condition needed in order to guarantee the independence of
physical observables of $\rho_1^{(j)}$ is 
\begin{eqnarray}
\beta^\rho = 0 \, .
\label{nec_cond}
\end{eqnarray}
Otherwise said, no spurious dependence on $\rho_1^{(j)}$ must be
generated via $\rho_1^{(j)}$-dependent 
normalization conditions for the coupling constant $g$
(equivalently, for the $\lambda^{(k)}_1$-counterterms).

\medskip
We remark that the above analysis relies on the fulfillment
of the ST identities in eq.(\ref{final_st}).
The non-symmetric counterterms in eq.(\ref{ym37}) are required in order
to ensure that eq.(\ref{final_st}) holds true at orders $n\geq 2$.
They must be included in order to guarantee the validity of 
eq.(\ref{ext4}).

\section{Conclusions}\label{sec4}

In this paper we have analyzed the restoration of the ST identities
for pure massless YM theory in the Landau gauge within the BPHZL
renormalization scheme and the Zimmermann-Lowenstein
prescription for handling massless propagators.
We have explicitly obtained the most general form of the action-like
part of the symmetric regularized action $\GG_s$, $s<1$, to all
orders in the loop expansion, and we have shown that non-symmetric
counterterms arise in $\GG^{(j)}_s$, $j\geq 2$, unless a special
choice of normalization conditions for $\GG^{(k)}_s$, $k<j$ is done.

We have verified that both UV and IR power-counting criteria
are fulfilled for $t^4 \GG_s$, thus guaranteeing the existence
of the physical limit $\GG = \lim_{s\rightarrow 1} \GG_s$
(absence of zero-mass singularities).

We have provided a cohomological interpretation of the IR power-counting
criterion, by noticing that it follows from the fact that the IR
regulator $m=M(s-1)$ enters into the classical action, together
with its BRST partner $\bar \rho$, only via a cohomologically
trivial term.

We have shown that the non-symmetric counterterms appearing
in $\GG^{(j)}_s$ at orders $j\geq 2$ do not vanish in the limit
$s\rightarrow 1$. We have analyzed the dependence of
the Green functions of physical observables on the coefficients
$\rho_1^{(k)}$ entering into the  parameterization of the non-symmetric
counterterms and we have discussed the associated Nielsen-like identities.

The proper inclusion of the non-symmetric counterterms
is strictly necessary in order to guarantee the fulfillment of the
ST identities at orders higher than one in the loop expansion,
for general lower-orders normalization conditions.

\medskip
Among the models with massless particles the most phenomenologically 
important ones are undoubtedly the SM and the MSSM.
The proof that the relevant ST identities can be restored
to all orders in perturbation theory has been given for the SM
in \cite{Kraus:1997bi,Grassi:1999nb} and for the MSSM in \cite{Hollik:2002mv}.
Still the explicit construction of the symmetric 1-PI generating functional
for the SM and the MSSM poses some problems,  due to
the lack of an invariant regularization scheme.
For the SM and especially for the MSSM the explicit recursive
evaluation of the ST breakings at the regularized level
and of the finite counterterms required to recover them
is significantly involved already at one loop %level 
and becomes
unfeasible at higher orders.

The direct restoration
of the ST identities, together with
the method for hadling massless particles analyzed in the present paper,
seems to be a promising tool in order to %
explicitly construct the symmetric 1-PI Green functions
for the SM and the MSSM, to all orders in the loop expansion.

\section*{Acknowledgments}

Useful discussions with R.~Ferrari and D.~Maison are gratefully
acknowledged. The Author wishes to thank the Theory Group
at the Milano University, where part of this work was completed,
for the kind hospitality.

\appendix
\section{The classical action}\label{appA}
The classical action of 
pure massless Yang-Mills theory in the Landau gauge with
an IR regulator $m$ introduced via the BRST doublet $(\bar \rho,m)$ 
is 
\begin{eqnarray}
&& \!\!\!\!\!\!\! \G^{(0)}  =  
\int d^4x \, \left \{ -\frac{1}{4g^2} G_{\mu\nu}^a G^{\mu\nu \, a}
- \bar \omega^a \partial_\mu
(D^\mu \omega)^a + B^a \partial A^a + A_\mu^{a*} (D^\mu \omega)^a 
\right . \nonumber \\
&& \left . ~~~~~~  - \omega^{* a} \frac{1}{2}f^{abc}
     \omega^b \omega^c + 
     \bar \omega^{a*} B^a 
   \right \} 
%\nonumber \\
%&&  
+ \int d^4x \, s \left (
\frac{1}{2}\bar \rho m (A_\mu^a)^2 + \bar \rho m \bar \omega^a \omega^a 
\right ) \nonumber \\
&& ~~~~~ = \int d^4x \, \left \{ -\frac{1}{4g^2} G_{\mu\nu}^a G^{\mu\nu \, a}
- \bar \omega^a \partial_\mu
(D^\mu \omega)^a + B^a \partial A^a+ A_\mu^{a*} (D^\mu \omega)^a
 \right . \nonumber \\
&& \left . ~~~~~~~~~~  - \omega^{* a} \frac{1}{2}f^{abc}
     \omega^b \omega^c + 
     \bar \omega^{a*} B^a 
+ 
 \frac{1}{2} m^2 (A_\mu^a)^2 + m^2 \bar \omega^a \omega^a
\right . \nonumber \\
&& \left . 
~~~~~~~~~~ - \bar \rho m A_\mu^a \partial^\mu \omega^a - \bar \rho m B^a \omega^a - \frac{1}{2} \bar \rho m \bar \omega^a f^{abc} \omega^b \omega^c \right \} \, .
\label{mod1}
\end{eqnarray}

$G_{\mu\nu}^a$ is the field strength
\begin{eqnarray}
G_{\mu\nu}^a = \partial_\mu A^a_\nu - \partial_\nu A_\mu^a + f^{abc} A_\mu^b A_\nu^c \, .
\label{app1}
\end{eqnarray}
$\omega^a$ is the ghost field, $\bar \omega^a$ the antighost field,
$B^a$ the associated Nakanishi-Lautrup multiplier field.

BRST transformations
\begin{eqnarray}
&& s A_\mu^a = (D_\mu \omega)^a = \partial_\mu \omega^a + f^{abc} A_\mu^b \omega^c \, , ~~~~ 
s \omega^a = -\frac{1}{2} f^{abc} \omega^b \omega^c \, , ~~~~ \nonumber \\
&& s \bar \omega^a = B^a \, , ~~~~ s B^a = 0 \, , \nonumber \\
&& s \bar \rho = m \, , ~~~~ s m = 0 \, .
\label{mod2}
\end{eqnarray}

\end{document}